# Model Dependent Analysis of $D_{(s)}^+ \to \eta^{(\prime)} l^+ \nu_l$ Decays in Beyond Standard Model

S. Mahata[1], M. Mandal, H. Mahapatra, S. Biswas and S. Sahoo[2]

Department of Physics, National Institute of Technology Durgapur
Durgapur 713209, West Bengal, India

[1] E-mail: supravatmahata3@gmail.com, [2] E-mail: sukadevsahoo@yahoo.com

**Abstract**

Motivated by the recent experimental results of branching fractions for $D_{(s)}^+ \to \eta^{(\prime)} \bar{l} \nu_l$ decays, which deviate from their SM predictions, we have investigated these decays in $W'$ model and scalar leptoquark model to find possible signatures of new physics (NP) in semileptonic charm decays induced by $c \to (s,d)\bar{l}\nu_l$ transitions. Using recent experimental results of branching fractions for semileptonic $D$ meson decays, new coupling parameters are predicted for the above NP models. Branching fraction, forward-backward asymmetry and lepton polarization asymmetry are studied taking the predicted NP coupling parameters. Results of branching fractions in scalar leptoquark model are found very close to the experimental results and exist around the range $1\sigma$ deviation. We have presented a comparative study of the NP models to check their sensitivity on these decays. We anticipate that further research on these decays will significantly support our findings.

## 1. Introduction

To explain our whole universe, searching of a fundamental theory becomes more prominent presently. Till now standard model (SM) is able to describe approximately 5% of our universe, so rest of the universe is unknown to us. It describes the three generations of quarks and leptons and their interactions (excluding gravity). Recent measurements on flavour physics and the existence of dark energy and dark matter indicate possible existence of the new physics (NP) beyond the SM. Therefore, the searching of NP has gained special attention to modify the particle physics. The new particles and their interactions with quarks and leptons should be the source of NP. Direct searches at high energy colliders and indirect searches for quantum effects in precise observables are generally the two methods used to look for new particles and interactions. The $b$ hadron decays provide a reliable platform to probe NP beyond the SM.

Recently, different experimental collaborations such as BaBar, Belle and LHCb have measured and updated the lepton flavour universality (LFU) ratios $R_{D^{(*)}}$ associated with flavour changing charged current (FCCC) $b \to c\bar{l}\nu_l$ transition. The reported deviations of $R_{D^{(*)}}$ from their corresponding SM prediction are around $3.3\sigma$ [1-3]. Another angular observable $P_5'$ [4] also displays hints for NP beyond the SM, which is the measurement of correlations among the trajectories of the decaying particle in $B_s \to K^{(*)}\mu^+\mu^-$ and the decay rate $\Gamma$ in $B_s \to \phi\mu^+\mu^-$.

There are so many NP models involving new particles, like non-universal $Z'$ models [5-10], leptoquark models [11-14] and models with charged Higgs [15-17] to explain such kind of anomalies. Presently, such kind of transitions have gained special attention both in experiments and in phenomenological study to explore NP. These tension between theory and experiment associated with $b$ hadron decays raise a question in our curious mind. Whether the similar inconsistency can be observed in charm decays induced by $c \to (s, d)\bar{l}\nu_l$ transitions! Most of the recent experimental measurements on the pure leptonic and semileptonic $D$ decays except $D_{(s)}^+ \to \eta^{(\prime)}\bar{l}\nu_l$ decays agree with the SM predictions. In the last few years, a lot of theoretical efforts have been done for searching NP contribution in $D$ decays [18 - 23] and charm decays provide a special opportunity to investigate flavour physics beyond the SM in the up-sector. From Table 1, it can be observed that the recent experimental results [24, 25] of branching fractions for the decays $D_{(s)}^+ \to \eta^{(\prime)}\bar{l}\nu_l$ depict tension with their corresponding SM predictions and hence, these decays provide opportunities to explore NP. Inspired by these results, the $D_{(s)}^+ \to \eta^{(\prime)}\bar{l}\nu_l$ decays are studied in $W'$ model [14, 26, 27] and scalar leptoquark (LQ) model [14] to explore NP effects. For theoretical study, the heavy-to-light form factors are key input parameters. Due to the nonperturbative nature, they can be parameterized using different nonperturbative methods such as quark models, QCD sum rules, lattice QCD (LQCD) [28-29] and light-cone sum rules (LCSR) [30]. Using LQCD approach [31], the SM predictions of branching fraction for $D \to \pi(K)l\nu_l$ decays are found to be good consistency with the experimental results [24] with large uncertainties arising from the CKM matrix elements. The LQCD approach is based on the first principles, it is more useful to get precise result for the $D \to P$ (pseudoscalar meson) transitions. The LCSR method utilizes the operator product expansion close to the light-cone, encapsulating all non-perturbative dynamics within the light-cone distribution amplitudes [32]. This approach is also suitable to parameterize heavy-to-light transition form factors. The parametric space of NP parameters have been constrained by taking precise prediction of $D$ mesons decays using LQCD and LCSR approaches. In the NP models, the $W'$ boson and leptoquark are the hypothetical beyond SM particles that couple a quark directly to a lepton unlike any particle within the SM. Both are the relics of grand unified theories, which were designed to integrate the various interactions in the SM at high energies. In this work, we have predicted new coupling parameters using recent experimental results of branching fractions given in Table 4 & 5. Using these new coupling parameters, branching fraction, forward-backward asymmetry and lepton polarization asymmetry of the $D_{(s)}^+ \to \eta^{(\prime)}\bar{l}\nu_l$ decays have been investigated to explore NP effects.

Table 1. Theoretical and experimental results of branching fractions and their corresponding deviation for the $D_{(s)}^+ \to \eta^{(\prime)}\bar{l}\nu_l$ decays.

| Decay | SM result [18] | Experiment [24] | $Pull_{SM}$ |
|---|---|---|---|
| $\mathcal{B}(D_s^+ \to \eta\mu^+\nu_\mu) \times 10^{-2}$ | $1.52 \pm 0.31$ | $2.4 \pm 0.5$ | $1.5\sigma$ |
| | | $2.215 \pm 0.051 \pm 0.052$ [25] | $2.2\sigma$ |
| $\mathcal{B}(D_s^+ \to \eta'\mu^+\nu_\mu) \times 10^{-3}$ | $5.64 \pm 1.10$ | $11.0 \pm 5.0$ | $1.05\sigma$ |
| | | $8.01 \pm 0.55 \pm 0.31$ [25] | $1.9\sigma$ |
| $\mathcal{B}(D_s^+ \to \eta e^+\nu_e) \times 10^{-2}$ | $1.55 \pm 0.33$ | $2.29 \pm 0.19$ | $2\sigma$ |
| $\mathcal{B}(D_s^+ \to \eta' e^+ \nu_e) \times 10^{-3}$ | $5.91 \pm 1.26$ | $7.4 \pm 1.4$ | $0.8\sigma$ |
| $\mathcal{B}(D^+ \to \eta \mu^+ \nu_\mu) \times 10^{-3}$ | $0.75 \pm 0.15$ | $1.04 \pm 0.11$ | $1.6\sigma$ |
| $\mathcal{B}(D^+ \to \eta' \mu^+ \nu_\mu) \times 10^{-4}$ | $1.06 \pm 0.20$ | - | - |
| $\mathcal{B}(D^+ \to \eta e^+ \nu_e) \times 10^{-3}$ | $0.76 \pm 0.16$ | $1.11 \pm 0.07$ | $2\sigma$ |
| $\mathcal{B}(D^+ \to \eta' e^+ \nu_e) \times 10^{-4}$ | $1.12 \pm 0.24$ | $2.0 \pm 0.4$ | $1.9\sigma$ |

The layout of this paper is as follows: In sec. 2, we go over the theoretical formulation for the effective field theory approach to study $D_{(s)}^+ \to \eta^{(\prime)} \bar{l} \nu_l$ decays. The formulation of $W'$ model and scalar LQ model are presented in sec. 3. In sec. 4, our numerical analysis is presented and a comparative study between our NP models is also conducted. We have concluded our findings in sec. 5.

## 2. Theoretical Framework

Considering all possible Lorentz structures and assuming only left-handed neutrinos, the general effective Lagrangian for the $c \to (s, d) \bar{l} \nu_l$ transitions can be written as [18, 19]

$$\mathcal{L}_{eff} = -\frac{4G_F}{\sqrt{2}} V_{cq}^* \left[ (1 + C_{V_L}^l) O_{V_L}^l + C_{V_R}^l O_{V_R}^l + C_{S_L}^l O_{S_L}^l + C_{S_R}^l O_{S_R}^l + C_T^l O_T^l \right] + h.c, \quad (1)$$

where $G_F$ is the Fermi constant and $V_{cq}$ is the CKM matrix element. The four-fermion operators are defined as

$$O_{V_L}^l = (\bar{q} \gamma^\mu P_L c)(\bar{\nu}_l \gamma_\mu P_L l), \qquad O_{V_R}^l = (\bar{q} \gamma^\mu P_R c)(\bar{\nu}_l \gamma_\mu P_L l),$$

$$O_{S_L}^l = (\bar{q} P_L c)(\bar{\nu}_l P_R l), \qquad O_{S_R}^l = (\bar{q} P_R c)(\bar{\nu}_l P_R l),$$

$$O_T^l = (\bar{q} \sigma^{\mu\nu} P_L c)(\bar{\nu}_l \sigma_{\mu\nu} P_R l), \quad (2)$$

where $P_{L(R)} = (1 \mp \gamma^5)/2$ and $C_i^l (i = V_L, V_R, S_L, S_R$ and $T)$ are the corresponding Wilson coefficients at the scale $\mu = m_c$, with $C_i^l = 0$ in SM.

Here, the hadronic transition is quantified using heavy-to-light form factors. These form factors, though rooted in nonperturbative dynamics, exhibit universal nature. Specifically, the hadronic matrix elements characterizing transitions from a $D$ meson to a pseudoscalar meson denoted as $P(P = \eta, \eta')$ can be expressed as

$$\langle P(p_2) | \bar{q} \gamma^\mu c | D(p_1) \rangle = f_+(q^2) \left[ (p_1 + p_2)^\mu - \frac{m_D^2 - m_P^2}{q^2} q^\mu \right] + f_0(q^2) \frac{m_D^2 - m_P^2}{q^2} q^\mu, \quad (3)$$

$$\langle P(p_2) | \bar{q} c | D(p_1) \rangle = \frac{q_\mu}{m_c - m_q} \langle P(p_2) | \bar{q} \gamma^\mu c | D(p_1) \rangle = \frac{m_D^2 - m_P^2}{m_c - m_q} f_0(q^2), \quad (4)$$

$$\langle P(p_2) | \bar{q} \sigma^{\mu\nu} c | D(p_1) \rangle = -i (p_1^\mu p_2^\nu - p_1^\nu p_2^\mu) \frac{2 f_T(q^2)}{m_D + m_P}, \quad (5)$$

where $q^\mu$ represents the momentum difference $(p_1 - p_2)^\mu$, while $f_+(q^2)$ and $f_0(q^2)$ correspond to a pair of QCD form factors that encapsulate the dynamics of strong interactions. These form factors satisfy the condition $f_+(0) = f_0(0)$, signifying a shared value at zero momentum transfer. At the present moment, the LQCD form factors connected to the above transitions remain inaccessible. In this work, we have used the form factors obtained from light cone sum rules (LCSR) [30], which are parameterized as

$$F^i(q^2) = \frac{F^i(0)}{1-a\frac{q^2}{m_D^2}+b\left(\frac{q^2}{m_D^2}\right)^2}, \qquad (6)$$

here, $F^i(q^2)$ has the flexibility to represent either of the form factors $f_+$ or $f_0$. Along with the CKM matrix element, the form factors describing the dynamic of strong interaction are important for the calculation of decay rate. The mesons $\eta$ and $\eta'$ stand out due to the significant involvement of the spectator quark in shaping the final state, unlike in the case of final-state hadrons $K$ and $\pi$. The $\eta$ and $\eta'$ mesons are the octet-singlet mixing state of $\eta - \eta'$ gluon [33, 34], this mixing parameter proves a deeper understanding of non-perturbative QCD confinement. The octet-singlet mixing angle $\theta$ is within the range $-10°$ to $-23°$ [35]. The mixtures of the flavour $SU(3)$ octet $\eta_8$ and singlet $\eta_0$ can be represented by

$$\begin{pmatrix}\eta\\\eta'\end{pmatrix} = \begin{pmatrix}\cos\theta & -\sin\theta\\\sin\theta & \cos\theta\end{pmatrix}\begin{pmatrix}\eta_8\\\eta_0\end{pmatrix}, \qquad (7)$$

where $\eta_8 = (u\bar{u} + d\bar{d} - 2s\bar{s})/\sqrt{6}$ and $\eta_0 = (u\bar{u} + d\bar{d} + s\bar{s})/\sqrt{3}$. For the quark-flavour basis $\eta_q = (u\bar{u} + d\bar{d})/\sqrt{2}$ and $\eta_s = s\bar{s}$, can be written as [36]

$$\begin{pmatrix}\eta\\\eta'\end{pmatrix} = \begin{pmatrix}\cos\phi & -\sin\phi\\\sin\phi & \cos\phi\end{pmatrix}\begin{pmatrix}\eta_q\\\eta_s\end{pmatrix}, \qquad (8)$$

where $\phi$ is the mixing angle between the state of $\eta_q$ and $\eta_s$. Due to the indefinite gluon content in the $\eta^{(\prime)}$, large uncertainty may arise. Hence the estimation of hadronic form factor is very important for the theoretical calculations. Using LQCD [37] and LCSR [38, 39], the $f_+^{\eta^{(\prime)}}(0)$ are estimated by considering the mixture of quarks and gluons for $\eta$ and $\eta'$ mesons. In the quark flavour basis, the $\eta - \eta'$ mixing angle is connected to the branching ratios of $D$ and $D_s$ by the relation $\cot^4\phi = \frac{\Gamma(D_s^+\to\eta' e^+\nu_e)/\Gamma(D_s^+\to\eta e^+\nu_e)}{\Gamma(D^+\to\eta' e^+\nu_e)/\Gamma(D^+\to\eta e^+\nu_e)}$, where a possible gluon component cancels out [40]. A supplementary constraint on the gluonium function in the $\eta^{(\prime)}$ mesons is provided by the determination of $\phi$, enhancing the knowledge of nonperturbative QCD dynamics and helps theoretical study of $D$ meson decays involving the $\eta^{(\prime)}$ mesons. The KLOE Collaboration [41] predicted the mixing angle $\phi = (39.7 \pm 0.7)°$ and $(41.5 \pm 0.3_{stat} \pm 0.7_{syst} \pm 0.6_{th})°$ with and without gluonium content for $\eta'$ respectively. Recently, BESIII Collaboration [42] has also established $\phi = (40.1 \pm 2.1_{stat} \pm 0.7_{syst})°$ from the measurement of dynamics of $D_s^+ \to \eta^{(\prime)} e^+\nu_e$ decays.

Table 2. Form factors for $D \to \eta_q$ and $D_s \to \eta_s$ [18, 30].

| Decay | $D \to \eta_q$ | | $D_s \to \eta_s$ | |
|---|---|---|---|---|
| Form factors | $f_+$ | $f_0$ | $f_+$ | $f_0$ |
| $F(0)$ | $0.56^{+0.06}_{-0.05}$ | $0.56^{+0.06}_{-0.05}$ | $0.61^{+0.06}_{-0.05}$ | $0.61^{+0.06}_{-0.05}$ |
| $a$ | $1.25^{-0.04}_{+0.05}$ | $0.65^{-0.01}_{+0.02}$ | $1.20^{-0.02}_{+0.03}$ | $0.64^{-0.01}_{+0.02}$ |
| $b$ | $0.42^{-0.06}_{+0.05}$ | $-0.22^{-0.03}_{+0.02}$ | $0.38^{-0.01}_{+0.01}$ | $-0.18^{+0.04}_{-0.03}$ |

The explanation of the kinematics involved in $D \to P \bar{l} \nu_l$ decay processes can be obtained through the helicity amplitudes. Within the framework of SM, the process where $c$ quark transitions to $s$ quark accompanied by the emission of an $l^+ \nu_l$ pair can be visualized as the $c$ quark transforming into an off-shell $W^{*+}$ boson. Subsequently, this off-shell $W^{*+}$ particle decays into an $l^+ \nu_l$ pair. From Eq. (1), the amplitude can be defined as [43-45]

$$\mathcal{M}(D \to P\bar{l}\nu_l) = \frac{G_F}{\sqrt{2}} V^*_{cq} \sum_k C_k(\mu) \langle P|\bar{c}\Gamma^k b|D\rangle \bar{u}_l \Gamma_k v_\nu , \qquad (9)$$

where $C_k(\mu)$ denotes the Wilson coefficient with values

$$C_k(\mu) = \begin{cases} 1 & \text{for SM,} \\ C_{V_{L,R}}, C_{S_{L,R}}, C_T & \text{for NP,} \end{cases}$$

where $\Gamma^k$ denotes the product of gamma matrices i.e., $\Gamma^k = \gamma^\mu(1 \pm \gamma_5)$ and $(1 \pm \gamma_5)$. The square of the matrix element can be expressed as the product of leptonic ($L_{\mu\nu}$) and hadronic ($H^{\mu\nu}$) tensors

$$|\mathcal{M}(D \to P\bar{l}\nu_l)|^2 = \frac{G_F^2}{2} |V_{cq}|^2 \sum_{i,j} C_{ij}(\mu)(L^{ij}_{\mu\nu}H^{\mu\nu,ij}) . \qquad (10)$$

Here the superscripts $i,j$ denote the combination of four operators $(V \mp A)$, $(S \mp P)$ in the effective lagrangian and the product of Wilson coefficients $C_i$ and $C_j$ is represented by $C_{ij}(\mu)$. In the following discussion, these superscripts have been omitted for convenience. The polarization vector of the off-shell particle $W^*(\epsilon^\mu(m))$ satisfies the orthonormality and completeness relations:

$$\epsilon^{*\mu}(m)\epsilon_\mu(n) = g_{mn}, \quad \sum_{m,n} \epsilon^{*\mu}(m)\epsilon^\nu(n)g_{mn} = g^{\mu\nu}, \qquad (11)$$

where $g_{mn} = diag(+,-,-,-)$ and $m,n = \pm, 0, t$ represent the transverse, longitudinal and time-like polarization components. Using above completeness relation, the product of $L_{\mu\nu}$ and $H^{\mu\nu}$ can be written as

$$L_{\mu\nu}H^{\mu\nu} = \sum_{m,m',n,n'} L(m,n)H(m',n')g_{mm'}g_{nn'}, \tag{12}$$

where $L(m,n) = L^{\mu\nu}\epsilon_\mu(m)\epsilon_\nu^*(n)$ and $H(m,n) = H^{\mu\nu}\epsilon_\mu^*(m)\epsilon_\nu(n)$ are the Lorentz invariant parameters and it is clear that their values are independent with respect to reference frame. In this context, $H(m,n)$ and $L(m,n)$ will be calculated in the $D$ meson rest frame and $l - \nu_l$ center-of-mass frame respectively.

The off-shell $W^{*+}$ encompasses four helicity states, characterized by $\lambda_W = \pm 1, 0$ (with $J = 1$) and $\lambda_W = 0$ (with $J = 0$). It is important to note that only the $W^{*+}$ possesses a timelike polarization, and the angular momenta are denoted by $J = 1\ or\ 0$ in the rest frame of the $W^*$ boson. To distinguish the two cases of $\lambda_W = 0$, we have assigned $\lambda_W = 0$ for $J = 1$ and $\lambda_W = t$ for $J = 0$. In the rest frame of the $D$ meson, the direction of motion of the $W^{*+}$ has been aligned along the $Z$- axis. In this context, the polarization vectors of the $W^{*+}$ are expressed as follows

$$\epsilon^\mu(\pm) = \frac{1}{2}(0,1,\pm i,0), \qquad \epsilon^\mu(0) = -\frac{1}{\sqrt{q^2}}(q_3,0,0,q_0), \qquad \epsilon^\mu(t) = -\frac{q^\mu}{\sqrt{q^2}}, \tag{13}$$

where $q^\mu$ denotes the four-momentum of the $W^{*+}$. The polarization vectors of the virtual $W^{*+}$ satisfy the above orthogonality and completeness relations.

Here, the total matrix can be separated into two distinct components: the lepton portion and the hadron portion. Interestingly neither of these components is inherently Lorentz invariant. However, upon introducing the completeness relations of the virtual $W^{*+}$ particle, both the hadron and lepton components transform into Lorentz invariant forms. This transformation allows us to adopt a specific coordinate system without compromising the integrity of the analysis. Consequently, we examine the hadron aspect in the initial state, particularly within the rest frame of $D$ meson, while simultaneously evaluating the lepton component in the rest frame of the virtual $W^{*+}$. The helicity amplitudes of $D \to PW^{*+}$ transitions are

$$H_{\lambda_W}^{PV}(q^2) = \epsilon_\mu^*(\lambda_W)\langle P(p_2)|\bar{q}\gamma^\mu c|D(p_1)\rangle, \tag{14}$$

$$H^{PS}(q^2) = \langle P(p_2)|\bar{q}c|D(p_1)\rangle, \tag{15}$$

$$H_{\lambda_W,\lambda_W'}^{PT}(q^2) = \epsilon_\mu^*(\lambda_W)\epsilon_\nu^*(\lambda_W')\langle P(p_2)|\bar{q}\sigma^{\mu\nu}c|D(p_1)\rangle. \tag{16}$$

The $D \to Pl^+\nu_l$ decays have contributions through only five helicity amplitudes, which are given by

$$H_0^{PV} = \frac{f_+\sqrt{Q_+Q_-}}{\sqrt{q^2}}, \quad H_t^{PV} = \frac{f_0 M_+ M_-}{\sqrt{q^2}}, \quad H^{PS} = \frac{f_0 M_+ M_-}{m_c - m_q},$$

$$H_{0,t}^{PT} = -\frac{f_T\sqrt{Q_+Q_-}}{M_+}, \quad H_{1,-1}^{PT} = -\frac{f_T\sqrt{Q_+Q_-}}{M_+}, \tag{17}$$

where $M_\pm = m_D \pm m_P$ and $Q_\pm = M_\pm^2 - q^2$. One can calculate the leptonic amplitudes directly and the explicit results are given in ref. [46].

Using the hadronic helicity amplitudes and the leptonic amplitudes, the two-fold differential angular decay distribution of $D \to Pl^+\nu_l$ decay can be expressed as

$$\frac{d^2\Gamma(D \to Pl^+\nu_l)}{dq^2 d\cos\theta_l} = \frac{G_F^2 |V_{cq}|^2 \sqrt{Q_+ Q_-}}{256\pi^3 m_D^3} \left(1 - \frac{m_l^2}{q^2}\right)^2 \left[q^2 A_1^P + \sqrt{q^2} m_l A_2^P + m_l^2 A_3^P\right], \qquad (18)$$

with

$$A_1^P = \left|C_{S_L} + C_{S_R}\right|^2 |H^{PS}|^2 + Re\left[(C_{S_L} + C_{S_R})C_T^*\right] H^{PS}(H_{0,t}^{PT} + H_{0,-1}^{PT})\cos\theta_l + 4|C_T|^2 \left|H_{0,t}^{PT} + H_{1,-1}^{PT}\right|^2 \cos^2\theta_l + \left|1 + C_{V_L} + C_{V_R}\right|^2 |H_0^{PV}|^2 \sin^2\theta_l, \qquad (19)$$

$$A_2^P = 2\{Re\left[(C_{S_L} + C_{S_R})(1 + C_{V_L} + C_{V_R})^*\right] H^{PS} H_t^{PV} - 2Re\left[C_T(1 + C_{V_L} + C_{V_R})^*\right] H_0^{PV}(H_{0,t}^{PT} + H_{1,-1}^{PT})\} - 2\{Re\left[(C_{S_L} + C_{S_R})(1 + C_{V_L} + C_{V_R})^*\right] H^{PS} H_0^{PV} - 2Re\left[C_T(1 + C_{V_L} + C_{V_R})^*\right] H_t^{PV}(H_{0,t}^{PT} + H_{1,-1}^{PT})\}\cos\theta_l, \qquad (20)$$

$$A_3^P = 4|C_T|^2 \left|H_{0,t}^{PT} + H_{1,-1}^{PT}\right|^2 \sin^2\theta_l + \left|1 + C_{V_L} + C_{V_R}\right|^2 (|H_0^{PV}|^2 \cos^2\theta_l - 2H_0^{PV} H_T^{PV} \cos\theta_l + |H_t^{PV}|^2), \qquad (21)$$

where $\theta_l$ is the angle between the charged lepton and opposite direction of the motion of the final meson in the virtual $W^{*+}$ rest frame. The forward-backward asymmetry is defined as

$$A_{FB}(q^2) = \frac{\int_0^1 d\cos\theta_l \frac{d^2\Gamma}{dq^2 d\cos\theta_l} - \int_{-1}^0 d\cos\theta_l \frac{d^2\Gamma}{dq^2 d\cos\theta_l}}{\int_0^1 d\cos\theta_l \frac{d^2\Gamma}{dq^2 d\cos\theta_l} + \int_{-1}^0 d\cos\theta_l \frac{d^2\Gamma}{dq^2 d\cos\theta_l}}, \qquad (22)$$

and the lepton polarization asymmetry is also expressed as

$$P_F^l(q^2) = \frac{\frac{d\Gamma(\lambda_l = \frac{1}{2})}{dq^2} - \frac{d\Gamma(\lambda_l = -\frac{1}{2})}{dq^2}}{\frac{d\Gamma(\lambda_l = \frac{1}{2})}{dq^2} + \frac{d\Gamma(\lambda_l = -\frac{1}{2})}{dq^2}}. \qquad (23)$$

## 3. NP Models

Motivated by various extended theoretical approaches of the SM for the prediction of hypothetical particles, recently, ATLAS and CMS collaborations [47-53] are involved in searching for heavy gauge bosons (like $W'$ boson, $Z'$ boson and leptoquarks) from the proton-proton collisions at $\sqrt{s} = 13$ TeV at LHC. By integrating theoretical principles with experimental findings, a range of phenomenological models for new physics (NP) have been constructed. These include the $W'$ model, the non-universal $Z'$ model, the Higgs-doublet model, the leptoquark model, the supersymmetry model, and among several others. To explore the effects of NP through the consideration of $W'$ boson and leptoquarks as intermediate particles, we have studied $c \to (s,d)\bar{l}\nu_l$ transitions in the framework of $W'$ model and scalar leptoquark model.

### 3.1 $W'$ Model

One of the beyond SM particle candidates is $W'$ boson which arises from the simplest extension of the electroweak gauge group $SU(2)_1 \times SU(2)_2 \times U(1)$. In this model, we have assumed that the exchanging of $W'$ boson with the SM $W$ boson causes the charged current transitions.

Now the most general Lorentz invariant Lagrangian to describe the couplings of $W'$ boson to quarks and leptons can be expressed as [14, 26, 27, 54, 55]

$$\mathcal{L}_{eff}^{W'} = \frac{W'_\mu}{\sqrt{2}} \left[ \bar{u}_i \left( \epsilon^L_{u_i d_j} P_L + \epsilon^R_{u_i d_j} P_R \right) \gamma^\mu d_j + \bar{l}_i \epsilon^L_{l_i v_j} \gamma^\mu P_L v_j \right] + h.c., \quad (24)$$

where $P_{L(R)} = (1 \mp \gamma_5)/2$ are the left (right)-handed chirality projectors; and the coefficients $\epsilon^L_{u_i d_j}, \epsilon^R_{u_i d_j}$ and $\epsilon^L_{l_i v_j}$ denote the dimensionless flavour-dependent coupling parameters with $u_i \in (u, c, t)$, $d_j \in (d, s, b)$ and $l_i, l_j \in (e, \mu, \tau)$. To simplify, the leptonic flavour dependent interactions ($i = j$) have been considered in this work. In the SM, $\epsilon^L_{u_i d_j} = g_L V_{u_i d_j}$ and $\epsilon^L_{l_i v_i} = g_L$ where $g_L$ is the $SU(2)_L$ gauge coupling constant and $V_{u_i d_j}$ is the corresponding CKM matrix element. In our model, the contribution of right-handed neutrino is ignored.

For $c \to q \bar{l} v_l$ transition, the SM effective Lagrangian can be defined as

$$-\mathcal{L}_{eff}(c \to q\bar{l}v_l)_{SM} = \frac{4 G_F V_{cq}}{\sqrt{2}} (\bar{q} \gamma^\mu P_L c)(\bar{v}_l \gamma_\mu P_L l). \quad (25)$$

From Eqs. (24) and (25), the total effective Lagrangian in the $W'$ model can be written as

$$-\mathcal{L}_{eff}(c \to q\bar{l}v_l)_{SM+W'} = \frac{4 G_F V_{cq}}{\sqrt{2}} \left[ (1 + C_{V_L})(\bar{q} \gamma^\mu P_L c)(\bar{v}_l \gamma_\mu P_L l) + C_{V_R} (\bar{q} \gamma^\mu P_R c)(\bar{v}_l \gamma_\mu P_L l) \right], \quad (26)$$

where $C_{V_L}$ and $C_{V_R}$ are the new Wilson coefficients corresponding to the left-handed and right-handed vector operator respectively. Comparing Eq. (26) with Eq. (1), these Wilson coefficients can be expressed as [14, 26]

$$C_{V_L} \equiv \frac{\sqrt{2}}{4 G_F V_{cq}} \frac{\epsilon^L_{qc} \epsilon^L_{v_l l}}{M^2_{W'}}$$

$$C_{V_R} \equiv \frac{\sqrt{2}}{4 G_F V_{cq}} \frac{\epsilon^R_{qc} \epsilon^L_{v_l l}}{M^2_{W'}}, \quad (27)$$

where $M_{W'}$ is the mass of the $W'$ boson, $\epsilon^{L,R}_{qc}$ and $\epsilon^L_{v_l l}$ are the effective flavour dependent $W'$ coupling parameters with quark and lepton respectively.

In order to study production cross section of the hypothetical particle at 95% confidence level (CL), for a wide range of masses and couplings, experimental searches have employed different models as benchmark scenarios. The coupling ranges have been set by several experimental studies to search for $W'$ boson, by considering the couplings of $W'$ boson to be the same as those of the SM $W$ boson. This particular model is known as the sequential SM (SSM). The $W'$ boson coupling strength $g_{W'}$ is expressed in term of the SM weak coupling strength $g_W = \frac{e}{\sin^2\theta_W} = 0.65$. Hence, their coupling ratio is $\frac{g_{W'}}{g_W} = 1$ in the SSM. The ATLAS and CMS collaboration have carried out search for this heavy boson in leptonic, semileptonic and hadronic final states. They have searched $W'$ boson in the $e\nu$ and $\mu\nu$ channels. ATLAS collaboration has set the lower limits of the $W'$ boson mass at 6.0 TeV and 5.1 TeV in the electron and muon channels in the SSM scenario [48]. Using ATLAS [56] and CMS [57] data, Greljo et al. [58] have fixed $W'$ coupling parameter for the quark transition $b \to c\tau\bar{\nu}_\tau$ with the $W'$ mass range [0.5-3.5] TeV. Gomez et al. [26] have also predicted the best-fitted value for this transition considering $M_{W'} = 1$ TeV. Here, we have considered the mass of $W'$ boson ~1 TeV and gauge coupling of $W'$ boson identical to that of SM $W$ boson, just like SSM for experimental studies. In addition, the couplings of this heavy boson to quark and lepton sectors are taken into account as having a non-universal nature in our $W'$ model.

Table 3. Values of the coupling parameters in $W'$ model.

| Current | Coupling Parameter | Mid value |
| --- | --- | --- |
| $c \to s\mu^+\nu_\mu$ | $\epsilon_{sc}^L \epsilon_{\nu_\mu \mu}^L$ | 1.01 |
| $c \to se^+\nu_e$ | $\epsilon_{sc}^L \epsilon_{\nu_e e}^L$ | 0.49 |
| $c \to d\mu^+\nu_\mu$ | $\epsilon_{dc}^L \epsilon_{\nu_\mu \mu}^L$ | 0.35 |
| $c \to de^+\nu_e$ | $\epsilon_{dc}^L \epsilon_{\nu_e e}^L$ | 0.34 |

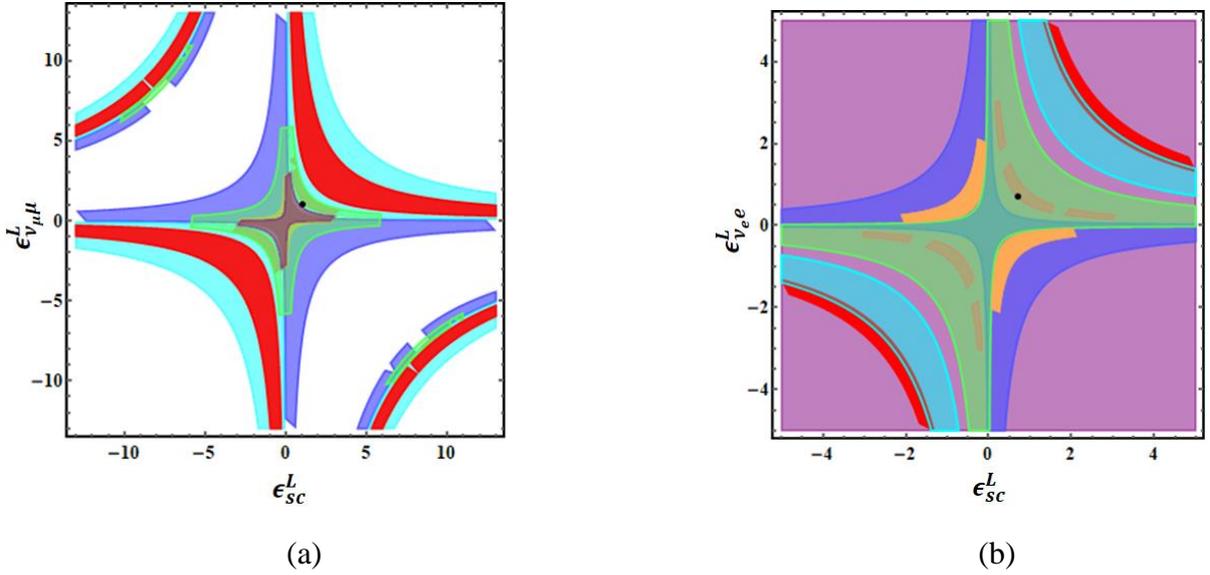

(a)  (b)

Fig 1. Allowed parameter space in the ($\epsilon_{sc}^L$, $\epsilon_{\nu_l l}^L$) plane for (a) $c \to s\mu^+\nu_\mu$ and (b) $c \to se^+\nu_e$ transitions. Here, colour notations are: Red ($D_s^+ \to \eta$), Cyan ($D_s^+ \to \eta'$), Brown ($D^+ \to \overline{K}^{*0}$), Blue ($D_s^+ \to \phi$), Green ($D^0 \to K^{*-}$), Orange ($D^+ \to \overline{K}^0$ and $D^0 \to K^-$) and Purple ($D_s^+ \to l^+\nu_l$). Black dot represents the fitted point of coupling parameters.

Table 4. The SM prediction of branching fractions for leptonic $D(D_s)$ decay and corresponding experimental results.

| Decay | SM Result [18] | Exp. Result [24] |
|---|---|---|
| $D^+ \to e^+ \nu_e$ | $(9.17 \pm 0.22) \times 10^{-9}$ | $< 8.8 \times 10^{-6}$ |
| $D^+ \to \mu^+ \nu_\mu$ | $(3.89 \pm 0.09) \times 10^{-4}$ | $(3.74 \pm 0.17) \times 10^{-4}$ |
| $D_s^+ \to e^+ \nu_e$ | $(1.24 \pm 0.02) \times 10^{-7}$ | $< 8.3 \times 10^{-5}$ |
| $D_s^+ \to \mu^+ \nu_\mu$ | $(5.28 \pm 0.08) \times 10^{-3}$ | $(5.43 \pm 0.15) \times 10^{-3}$ |

Table 5. The SM prediction of branching fractions for semileptonic $D(D_s)$ decays calculated using LQCD [28, 29] and LCSR [30] and corresponding experimental results [24].

| $c \to s \mu^+ \nu_\mu$ | | |
|---|---|---|
| Decay | SM Result [18] | Exp. Result [24] |
| $D^0 \to K^- \mu^+ \nu_\mu$ | $(3.40 \pm 0.22) \times 10^{-2}$ | $(3.41 \pm 0.04) \times 10^{-2}$ |
| $D^+ \to \bar{K}^0 \mu^+ \nu_\mu$ | $(8.69 \pm 0.57) \times 10^{-2}$ | $(8.76 \pm 0.19) \times 10^{-2}$ |
| $D^0 \to K^{*-} \mu^+ \nu_\mu$ | $(1.81 \pm 0.16) \times 10^{-2}$ | $(1.89 \pm 0.24) \times 10^{-2}$ |
| $D^+ \to \bar{K}^{*0} \mu^+ \nu_\mu$ | $(4.71 \pm 0.42) \times 10^{-2}$ | $(5.27 \pm 0.15) \times 10^{-2}$ |
| $D_s^+ \to \phi \mu^+ \nu_\mu$ | $(2.33 \pm 0.40) \times 10^{-2}$ | $(1.90 \pm 0.50) \times 10^{-2}$ |
| $D_s^+ \to \eta \mu^+ \nu_\mu$ | $(1.52 \pm 0.31) \times 10^{-2}$ | $(2.4 \pm 0.5) \times 10^{-2}$ |
| $D_s^+ \to \eta' \mu^+ \nu_\mu$ | $(5.64 \pm 1.10) \times 10^{-3}$ | $(11.0 \pm 5.0) \times 10^{-3}$ |
| $c \to s e^+ \nu_e$ | | |
| Decay | SM Result [18] | Exp. Result [24] |
| $D^0 \to K^- e^+ \nu_e$ | $(3.49 \pm 0.23) \times 10^{-2}$ | $(3.542 \pm 0.0035) \times 10^{-2}$ |
| $D^+ \to \bar{K}^0 e^+ \nu_e$ | $(8.92 \pm 0.59) \times 10^{-2}$ | $(8.73 \pm 0.10) \times 10^{-2}$ |
| $D^0 \to K^{*-} e^+ \nu_e$ | $(1.92 \pm 0.17) \times 10^{-2}$ | $(2.15 \pm 0.16) \times 10^{-2}$ |
| $D^+ \to \bar{K}^{*0} e^+ \nu_e$ | $(4.98 \pm 0.45) \times 10^{-2}$ | $(5.40 \pm 0.10) \times 10^{-2}$ |
| $D_s^+ \to \phi e^+ \nu_e$ | $(2.46 \pm 0.42) \times 10^{-2}$ | $(2.39 \pm 0.16) \times 10^{-2}$ |
| $D_s^+ \to \eta e^+ \nu_e$ | $(1.55 \pm 0.33) \times 10^{-2}$ | $(2.29 \pm 0.19) \times 10^{-2}$ |
| $D_s^+ \to \eta' e^+ \nu_e$ | $(5.91 \pm 1.26) \times 10^{-3}$ | $(7.4 \pm 1.4) \times 10^{-3}$ |
| $c \to d \mu^+ \nu_\mu$ | | |
| Decay | SM Result [18] | Exp. Result [24] |
| $D^0 \to \pi^- \mu^+ \nu_\mu$ | $(2.60 \pm 0.31) \times 10^{-3}$ | $(2.67 \pm 0.12) \times 10^{-3}$ |
| $D^+ \to \pi^0 \mu^+ \nu_\mu$ | $(3.37 \pm 0.40) \times 10^{-3}$ | $(3.50 \pm 0.15) \times 10^{-3}$ |
| $D^0 \to \rho^- \mu^+ \nu_\mu$ | $(1.65 \pm 0.23) \times 10^{-3}$ | $(1.35 \pm 0.13) \times 10^{-3}$ |
| $D^+ \to \rho^0 \mu^+ \nu_\mu$ | $(2.14 \pm 0.30) \times 10^{-3}$ | $(2.4 \pm 0.4) \times 10^{-3}$ |
| $D^+ \to \omega^0 \mu^+ \nu_\mu$ | $(1.82 \pm 0.26) \times 10^{-3}$ | $(1.77 \pm 0.21) \times 10^{-3}$ |
| $D^+ \to \eta \mu^+ \nu_\mu$ | $(0.75 \pm 0.15) \times 10^{-3}$ | $(1.04 \pm 0.11) \times 10^{-3}$ |
| $c \to d e^+ \nu_e$ | | |
| Decay | SM Result [18] | Exp. Result [24] |
| $D^0 \to \pi^- e^+ \nu_e$ | $(2.63 \pm 0.32) \times 10^{-3}$ | $(2.91 \pm 0.04) \times 10^{-3}$ |
| $D^+ \to \pi^0 e^+ \nu_e$ | $(3.41 \pm 0.41) \times 10^{-3}$ | $(3.72 \pm 0.17) \times 10^{-3}$ |
| $D_s^+ \to K^{*0} e^+ \nu_e$ | $(2.33 \pm 0.34) \times 10^{-3}$ | $(2.15 \pm 0.28) \times 10^{-3}$ |
| $D^0 \to \rho^- e^+ \nu_e$ | $(1.74 \pm 0.25) \times 10^{-3}$ | $(1.50 \pm 0.12) \times 10^{-3}$ |
| $D^+ \to \rho^0 e^+ \nu_e$ | $(2.25 \pm 0.32) \times 10^{-3}$ | $(2.18^{+0.17}_{-0.25}) \times 10^{-3}$ |
| $D^+ \to \omega^0 e^+ \nu_e$ | $(1.91 \pm 0.27) \times 10^{-3}$ | $(1.69 \pm 0.11) \times 10^{-3}$ |
| $D^+ \to \eta e^+ \nu_e$ | $(0.76 \pm 0.16) \times 10^{-3}$ | $(1.11 \pm 0.07) \times 10^{-3}$ |
| $D^+ \to \eta' e^+ \nu_e$ | $(1.12 \pm 0.24) \times 10^{-4}$ | $(2.0 \pm 0.4) \times 10^{-4}$ |

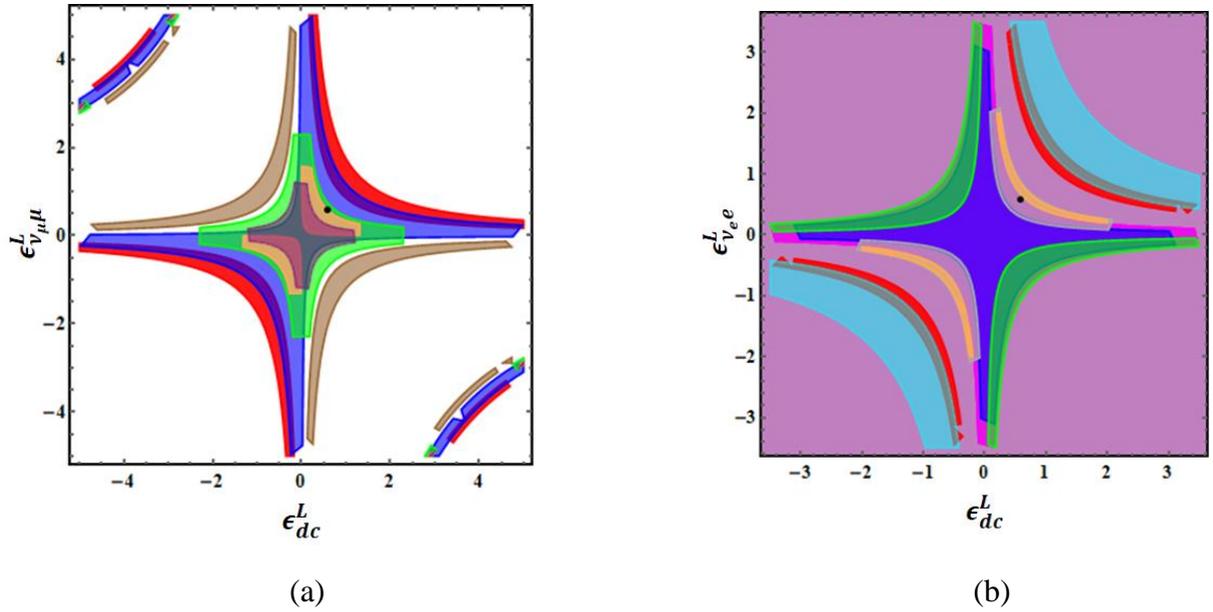

(a)                (b)

Fig 2. Allowed parameter space in the $(\epsilon_{dc}^L, \epsilon_{\nu_l l}^L)$ plane for (a) $c \to d\mu^+ \nu_\mu$ and (b) $c \to d e^+ \nu_e$ transitions. Here, colour notations are: Red $(D^+ \to \eta)$, Cyan $(D^+ \to \eta')$, Blue $(D^+ \to \rho^0)$, Green $(D^+ \to \omega^0)$, Brown $(D^0 \to \rho^-)$, Magenta $(D_s^+ \to K^{*0})$, Gray $(D^+ \to \pi^0)$, Orange $(D^0 \to \pi^-)$ and Purple $(D^+ \to l^+ \nu_l)$. Black dot represents the fitted point of coupling parameters.

In this work, we have adopted this model to study $c \to (s,d)\bar{l}\nu_l$ transitions for the first time. Only the effect of left-handed $W'$ boson has been considered in our model i.e., $C_{V_R} = 0$, right-handed vector operator $(\bar{q}\gamma^\mu P_R c)(\bar{v}_l \gamma_\mu P_L l)$ does not contribute to the LFU violation at leading order [59]. New coupling parameters $\epsilon_{qc}^L$ and $\epsilon_{v_l l}^L$ have been fixed from the region plots (Figs. 1 & 2) using recent experimental results of branching fractions associated with the $c \to (s,d)\bar{l}v_l$ transitions which are summarized in Table 3.

### 3.2 Scalar Leptoquark Model

In beyond SM, quarks can transform into leptons and vice versa through the intermediate hypothetical particles called leptoquarks (LQ). It is possible by grand unification theories (GUT) in which matter and forces are united. So, the transformation between quark and lepton has a higher order symmetry multiplet. Recently, searching of LQ in theory and experiments become more economical to extend the SM [60-63]. Leptoquarks have both spin zero ($s = 0$) – scalar leptoquarks and one ($s = 1$) – vector leptoquarks, they also carry colour and fractional electric charge. Experimental searches for LQs at colliders have been carried out by considering the same couplings to all SM fermions or enhanced couplings to second- or third-generation fermions. From the $pp \to \tau v$ process, which is mediated by exchange of LQ, lower limits on the masses of LQ are set at $205/515/5900$ GeV for the best fit LH / best-fit LH+RH / democratic scenarios by CMS collaboration [50]. From the significant part of the LQ parameter space, CMS have probed for LQ with masses up to 10 TeV and LQ couplings up to 8. The LQ mass has been set in the range $0.98 - 1.73$ TeV, depending on the LQ spin and its coupling to a lepton and a quark [51]. We have taken the LQ mass ~1 TeV in our scalar LQ model and the coupling to the quarks and leptons have been considered as non-universal. Considering mass $M_\phi \sim 1$ TeV and $O(1)$ coupling, a NP model is developed in the Ref. [63] which can explain $R_{D^*}$ anomalies. To describe the transformation between quark and lepton through LQ as mediator, the Lagrangian can be defined as [14, 64]

$$\mathcal{L}_{int}^\phi \supset \bar{Q}_L^c \lambda^L i\tau_2 L \phi^* + \bar{u}_R^c \lambda^R l_R \phi^* + h.c., \quad (28)$$

where $Q_L$, $L$ present the left-handed quark and lepton doublet, and $u_R$, $l_R$ present the right-handed up-type quark and lepton singlet respectively. The Yukawa coupling matrices in flavour space are denoted by $\lambda^{L,R}$. The basic key feature of the model is that the $c \to q \bar{l} v_l$ transition occurs through the vector ($O_{V_L}$), scalar ($O_{S_L}$) and tensor ($O_T$) currents by intermediate scalar LQ $\phi$ and their corresponding Wilson coefficients are defined as [14]

$$C_{V_L}(M_\phi) = \frac{\lambda_{cv_l}^L \lambda_{ql}^{L*}}{4\sqrt{2} G_F V_{cq} M_\phi^2}, \qquad C_{S_L}(M_\phi) = -\frac{\lambda_{cv_l}^L \lambda_{ql}^{R*}}{4\sqrt{2} G_F V_{cq} M_\phi^2}, \qquad C_T(M_\phi) = -\frac{1}{4} S_L(M_\phi). \quad (29)$$

Using recent experimental results of branching fractions given in Table 4 & 5 and taking $M_\phi = 1$ TeV, the new Yukawa couplings are predicted from the region plots given in the Figs. 3 and 4. The results are summarized in Table 6. We have taken 10% uncertainty of the coupling parameters for our study.

Table 6. Values of the scalar LQ coupling parameters.

| Current | Coupling Parameter | Mid value |
|---|---|---|
| $c \to s\mu^+ v_\mu$ | $\lambda_{cv_\mu}^L \lambda_{s\mu}^{L*}$ | $4.30$ |
| | $\lambda_{cv_\mu}^L \lambda_{s\mu}^{R*}$ | $-0.47$ |
| $c \to se^+\nu_e$ | $\lambda^L_{c\nu_e}\lambda^{L*}_{se}$ | $1.63$ |
| | $\lambda^L_{c\nu_e}\lambda^{R*}_{se}$ | $-0.44$ |
| $c \to d\mu^+\nu_\mu$ | $\lambda^L_{c\nu_\mu}\lambda^{L*}_{d\mu}$ | $0.78$ |
| | $\lambda^L_{c\nu_\mu}\lambda^{R*}_{d\mu}$ | $-0.05$ |
| $c \to de^+\nu_e$ | $\lambda^L_{c\nu_e}\lambda^{L*}_{de}$ | $1.0$ |
| | $\lambda^L_{c\nu_e}\lambda^{R*}_{de}$ | $-0.18$ |

## 4. Numerical Analysis

In this paper, we study $D^+_{(s)} \to \eta^{(\prime)}\bar{l}\nu_l$ decays involving $c \to (s,d)\bar{l}\nu_l$ transitions in the above two NP models. At first, we have structured these two models and then all the necessary coupling parameters have predicted from region plots using recent experimental results of branching fractions summarized in Table 4 & 5. Predicted coupling parameters are given in Table 3 and 6. The hadronic form factors are one of the key input parameters for this theoretical study. Till now LQCD results are unavailable, so we have used the results calculated by LCSR in our calculation which are presented in Table 2. The values of other input parameters are also summarized in the appendix.

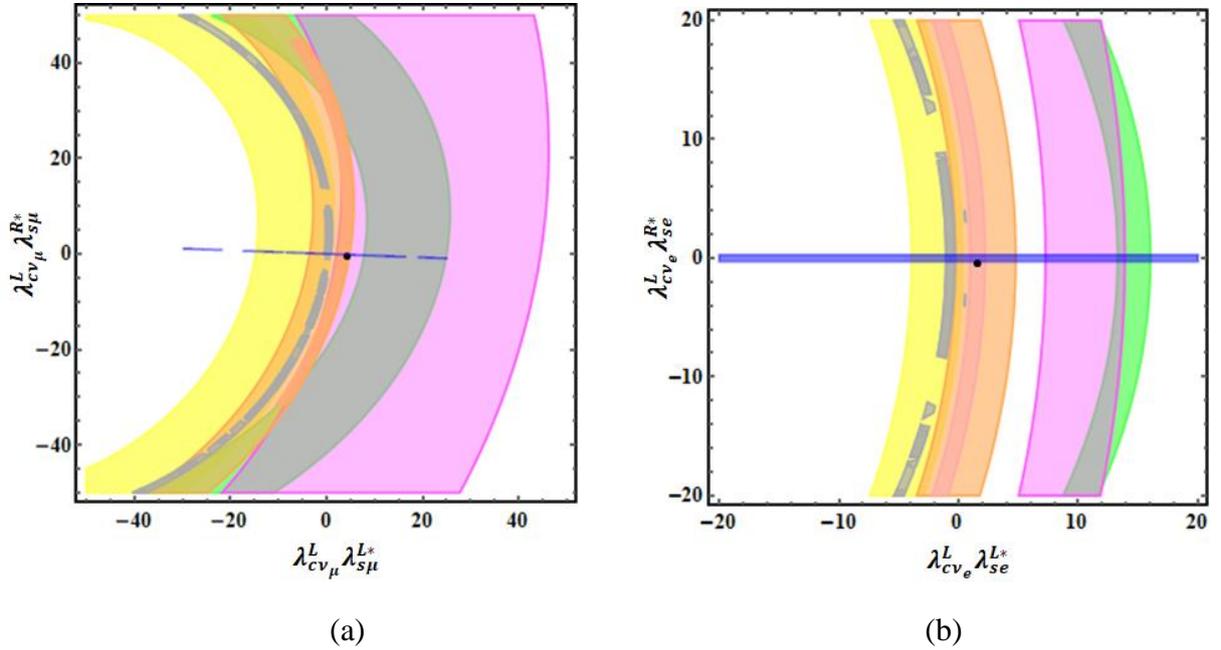

(a)            (b)

Fig 3. Allowed parameter space in the $(\lambda^L_{c\nu_l}\lambda^{L*}_{sl}, \lambda^L_{c\nu_l}\lambda^{R*}_{sl})$ plane for (a) $c \to s\mu^+\nu_\mu$ and (b) $c \to se^+\nu_e$ transitions. Here, colour notations are: Green ($D^+_s \to \eta$), Magenta ($D^+_s \to \eta'$), Yellow ($D^+_s \to \phi$), Pink ($D^+ \to \bar{K}^{*0}$), Orange ($D^0 \to K^{*-}$), Grey ($D^0 \to K^-$ and $D^+ \to \bar{K}^0$) and Blue ($D^+_s \to l^+\nu_l$). Black dot represents the fitted point of coupling parameters.

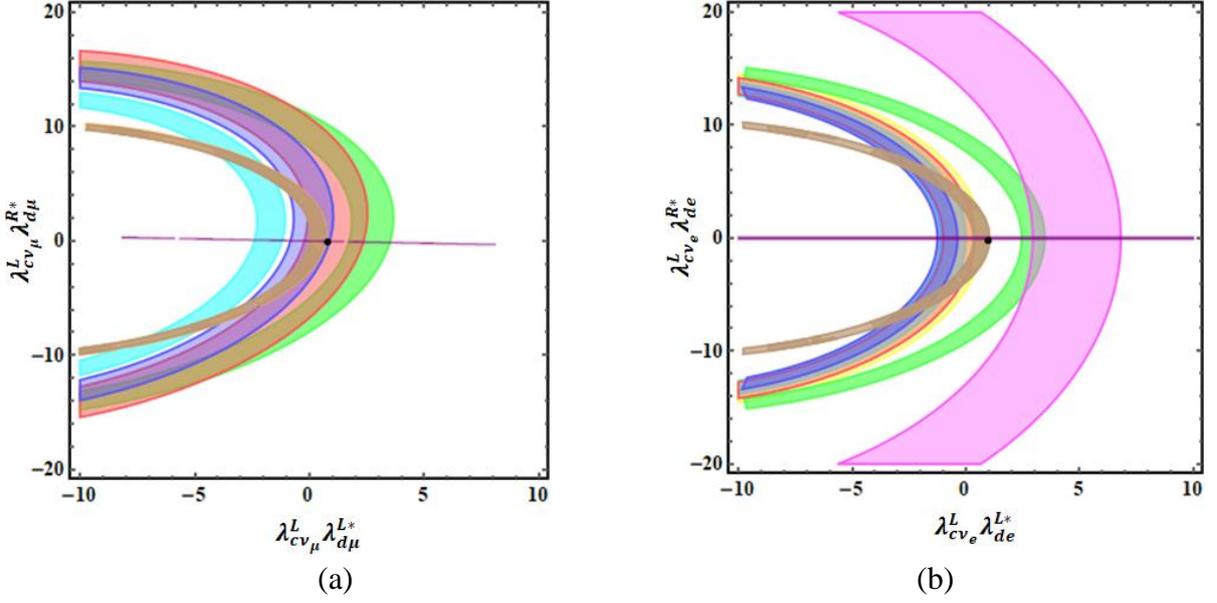

Fig 4. Allowed parameter space in the ($\lambda^L_{cv_l}\lambda^{L*}_{dl}$, $\lambda^L_{cv_l}\lambda^{R*}_{dl}$) plane for (a) $c \to d\mu^+\nu_\mu$ and (b) $c \to de^+\nu_e$ transitions. Here, colour notations are: Green ($D^+ \to \eta$), Magenta ($D^+ \to \eta'$), Blue ($D^+ \to \omega$), Cyan ($D^0 \to \rho^-$), Red ($D^+ \to \rho^0$), Yellow ($D^+_s \to K^{0-}$), Brown ($D^0 \to \pi^-$ and $D^+ \to \pi^0$) and Purple ($D^+ \to l^+\nu_l$). Black dot represents the fitted point of coupling parameters.

Using $W'$ model, we have mainly investigated differential branching fraction for the above decays. The $q^2$ dependence of differential branching fractions have been depicted in Figs 5 and 6. The significant contributions of $W'$ boson has been observed in all decays except $D^+_s \to \eta^{(\prime)}e^+\nu_e$ decays. All the results of branching fractions are given in Table 7 and results using this model found around 1.5 $\sigma$ range of the experimental results. Basically $W'$ model has NP contribution through the left-handed vector Willson coefficient ($C_{V_L}$) corrosponding to the vector operator ($O_{V_L}$). The $C_{V_L}$ cancels out in the other two observables forward-backward asymmetry $A_{FB}(q^2)$ and lepton polarization asymmetry $P^l_F(q^2)$. Therefore, this model is unable to provide NP in these decays and it is only suitable to study branching fraction.

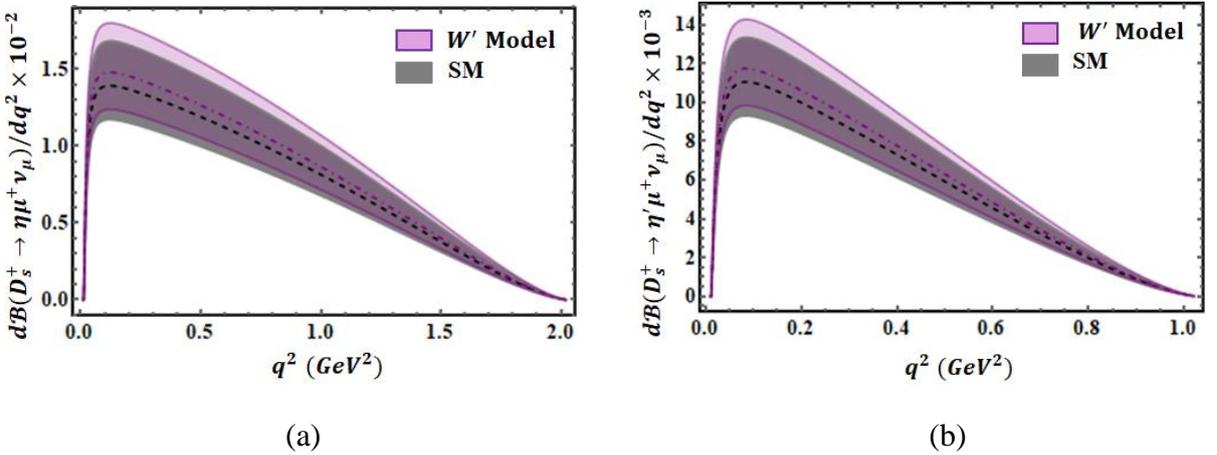

(a)    (b)

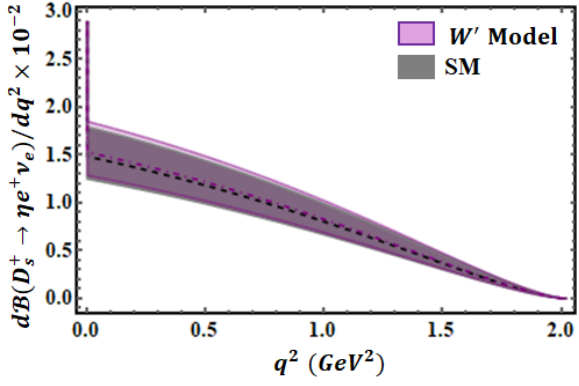
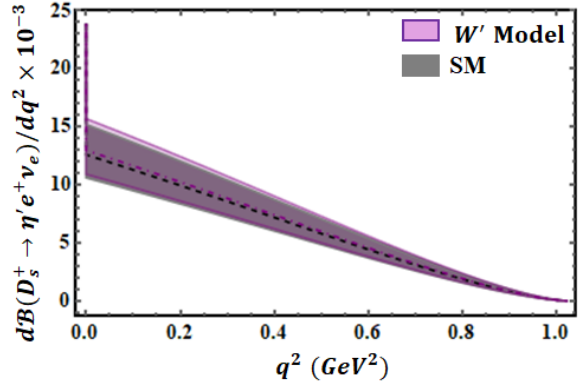

(c)  (d)

Fig 5. $q^2$ dependence of branching fractions for $D_s^+ \to \eta^{(\prime)} \bar{l} \nu_l$ in SM and $W'$ model. Here, black dashed and purple dash-doted lines present the central variation in SM and $W'$ model.

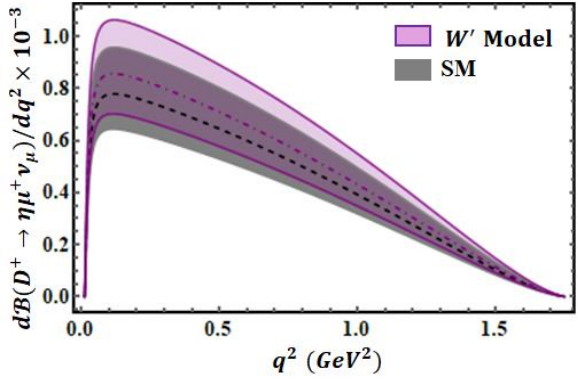
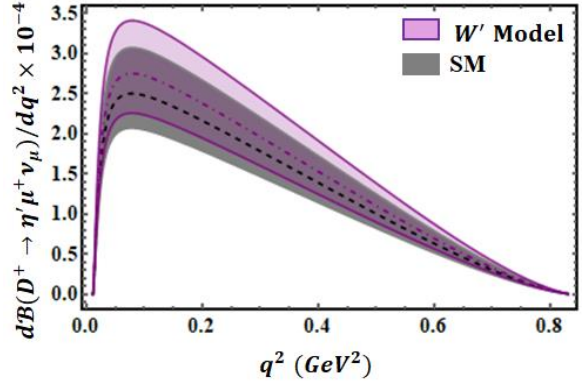

(a)  (b)

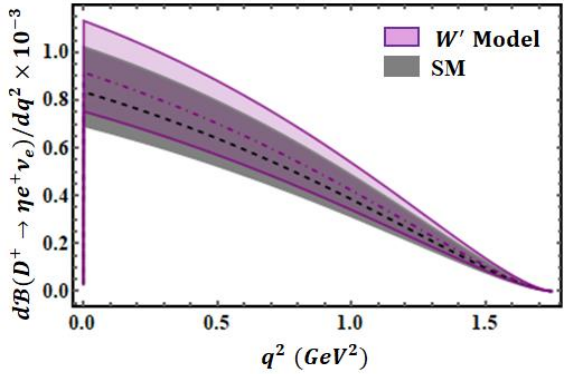
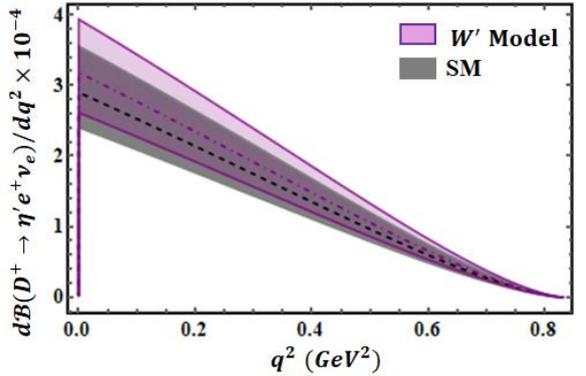

(c)  (d)

Fig 6. $q^2$ dependence of branching fractions for $D^+ \to \eta^{(\prime)} \bar{l} \nu_l$ in SM and $W'$ model. Here, black dashed and purple dash-doted lines present the central variation in SM and $W'$ model.

Next, we have focused on scalar LQ model. It provides NP effect through the vector, scalar and tensor operators. Therefore, it is capable to study NP dependent decay observables for finding possible NP signal. In our work, differential branching fraction, forward-backward asymmetry and lepton polarization asymmetry have been investigated in this model. In the Figs. 7 and 8, $q^2$ dependence of branching fractions have been shown. The significant effects of scalar LQ model on diffential branching fraction have been observed in all decays. Using this model, our predicted values of branching fractions and their corresponding pulls from the experimental results are given in Table 7. All results have been founded round the $1\sigma$ range of the experimental results excludeing $D_s^+ \to \eta e^+ \nu_e$ and $D^+ \to \eta' e^+ \nu_e$ decays. Hence, this model successfully explains all of the experimental findings better than the $W'$ model. Any contribution of scalar LQ is not observed for the forward-backward asymmetry and lepton polarization asymmetry. In the NP framework, these two decay observables mainly depend on the NP Wilson coefficient $C_T$ and this equals to the one forth of $C_{S_L}$. By putting all the values in Eq (29), we get $C_{S_L} = 0.7 \times 10^{-2}$ and $C_T = 0.2 \times 10^{-2}$ for the $c \to s \mu^+ \nu_\mu$ transition. Due to this very small value of $C_T$, noticeable effect of scalar LQ on these obsevable has not been found in Scalar LQ model.

Table 7. Values of the branching fractions in $W'$ model and scalar LQ model and their corresponding pull from the experimental results.

| Decay | $W'$ Model | Scalar LQ Model | Experiment [24] | $Pull_{SM}$ | $Pull_{W'}$ | $Pull_{LQ}$ |
|---|---|---|---|---|---|---|
| $\mathcal{B}(D_s^+ \to \eta \mu^+ \nu_\mu) \times 10^{-2}$ | $1.61 \pm 0.37$ | $1.73 \pm 0.41$ | $2.4 \pm 0.5$ | $1.5\sigma$ | $1.3\sigma$ | $1.0\sigma$ |
| | | | $2.215 \pm 0.051 \pm 0.052$ [25] | $2.2\sigma$ | $1.6\sigma$ | $1.1\sigma$ |
| $\mathcal{B}(D_s^+ \to \eta' \mu^+ \nu_\mu) \times 10^{-3}$ | $6.00 \pm 1.31$ | $6.42 \pm 1.47$ | $11.0 \pm 5.0$ | $1.05\sigma$ | $0.9\sigma$ | $0.9\sigma$ |
| | | | $8.01 \pm 0.55 \pm 0.31$ [25] | $1.9\sigma$ | $1.3\sigma$ | $0.8\sigma$ |
| $\mathcal{B}(D_s^+ \to \eta e^+ \nu_e) \times 10^{-2}$ | $1.59 \pm 0.36$ | $1.62 \pm 0.38$ | $2.29 \pm 0.19$ | $2\sigma$ | $1.7\sigma$ | $1.5\sigma$ |
| $\mathcal{B}(D_s^+ \to \eta' e^+ \nu_e) \times 10^{-3}$ | $6.09 \pm 1.32$ | $6.21 \pm 1.36$ | $7.4 \pm 1.4$ | $0.8\sigma$ | $0.7\sigma$ | $0.6\sigma$ |
| $\mathcal{B}(D^+ \to \eta \mu^+ \nu_\mu) \times 10^{-3}$ | $0.82 \pm 0.21$ | $0.83 \pm 0.21$ | $1.04 \pm 0.11$ | $1.6\sigma$ | $0.9\sigma$ | $0.9\sigma$ |
| $\mathcal{B}(D^+ \to \eta' \mu^+ \nu_\mu) \times 10^{-4}$ | $1.15 \pm 0.28$ | $1.16 \pm 0.29$ | - | - | - | - |
| $\mathcal{B}(D^+ \to \eta e^+ \nu_e) \times 10^{-3}$ | $0.83 \pm 0.22$ | $0.87 \pm 0.23$ | $1.11 \pm 0.07$ | $2\sigma$ | $1.2\sigma$ | $0.9\sigma$ |
| $\mathcal{B}(D^+ \to \eta' e^+ \nu_e) \times 10^{-4}$ | $1.21 \pm 0.30$ | $1.26 \pm 0.31$ | $2.0 \pm 0.4$ | $1.9\sigma$ | $1.6\sigma$ | $1.5\sigma$ |

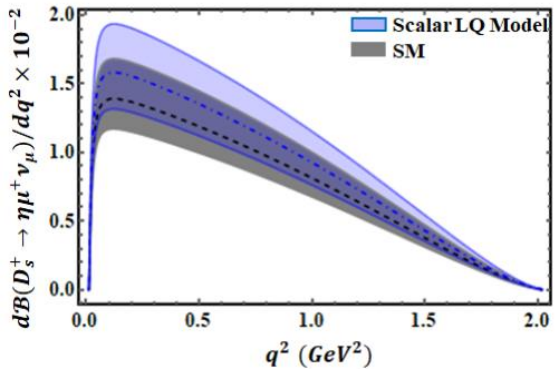
(a)

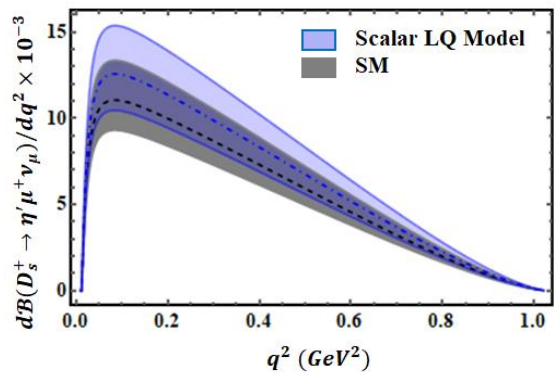
(b)

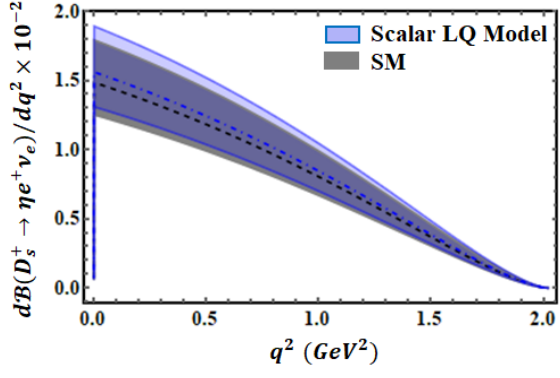
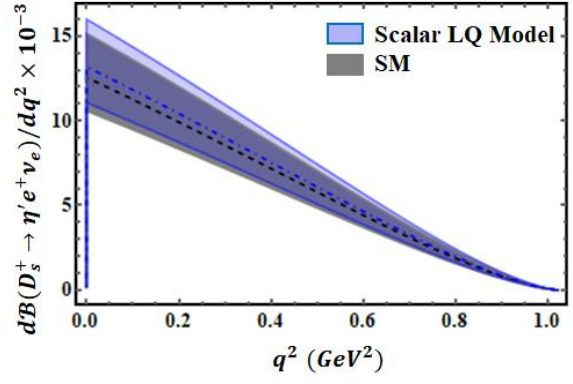

(c)                                              (d)

Fig 7. $q^2$ dependence of branching fractions for $D_s^+ \to \eta^{(\prime)} \bar{l} \nu_l$ decays in SM and scalar LQ model. Here, black dashed and blue dash-doted lines present the central variation in SM and scalar LQ model.

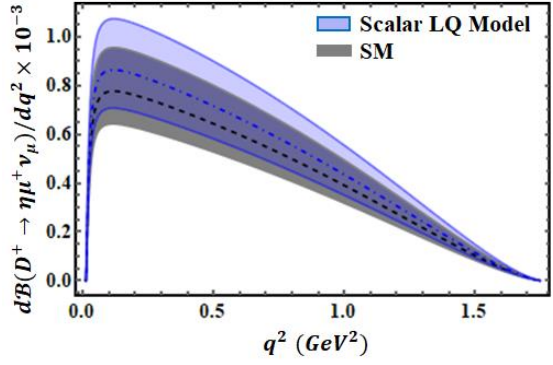
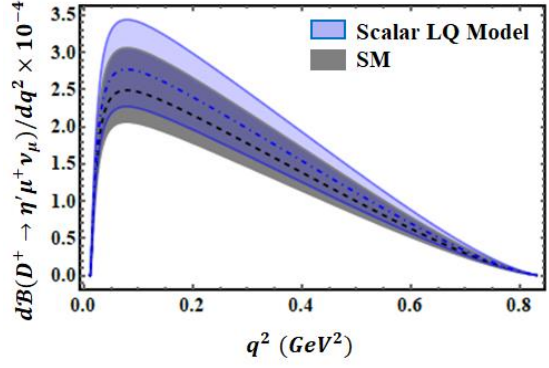

(a)                                              (b)

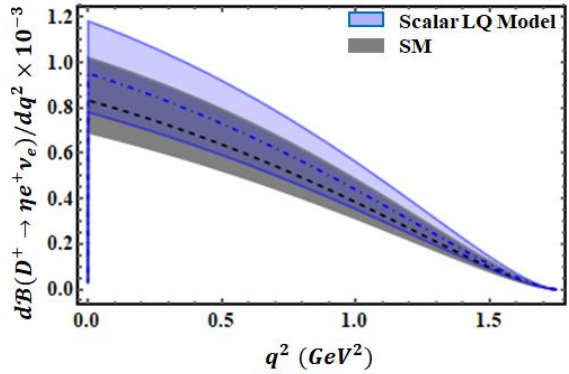
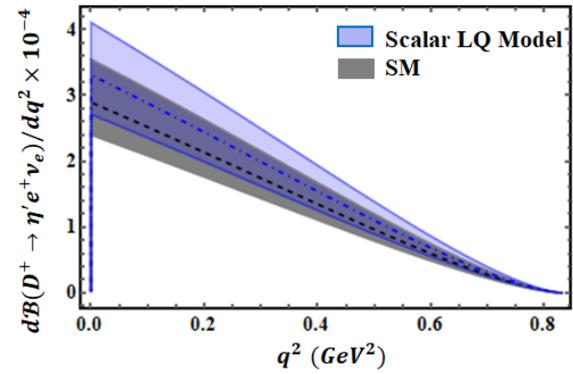

(c)                                              (d)

Fig 8. $q^2$ dependence of branching fractions for $D^+ \to \eta^{(\prime)} \bar{l} \nu_l$ decays in SM and Scalar LQ model. Here black dashed and blue dash-doted lines present the central variation in SM and scalar LQ model.

    Finally, we have compared the $W'$ model and scalar LQ model together with the experiment and the SM in 2D ($\mathcal{B}(D_{(s)}^+ \to \eta l^+ \nu_l)$, $\mathcal{B}(D_{(s)}^+ \to \eta' l^+ \nu_l)$) plane. These comparisons

are shown in Fig 9. For $D_s^+ \to \eta^{(\prime)}\mu^+\nu_\mu$ decays (Fig. 9(a)), the branching fraction results from the scalar LQ model are substantially closer to the experimental results. The results using $W'$ model are found within the SM and experimental values. Moreover, our results are also found to be more consistent with the recent results determined by the BESIII collaboration [25]. In $D_s^+ \to \eta^{(\prime)}e^+\nu_e$ decay (Fig. 9(b)), the values in LQ model have been obseved to shift toword the experiment results. Using $W'$ model, we observed slight deviation from the SM for these decay channels. Experimentally $D^+ \to \eta'\mu^+\nu_\mu$ decay is not studied yet, we have compared our NP results with the SM results for $D^+ \to \eta^{(\prime)}\mu^+\nu_\mu$ decays (Fig. 9(c)). The same amount of deviation obtained in both models. For $D^+ \to \eta^{(\prime)}e^+\nu_e$ decays (Fig. 9(d)), we have found significant deviation compare to the $W'$ model in Scalar LQ model. From this comparative study, it is clear that the scalar LQ model is more reliable than the $W'$ model to study $c \to (s,d)\bar{l}\nu_l$ transitions.

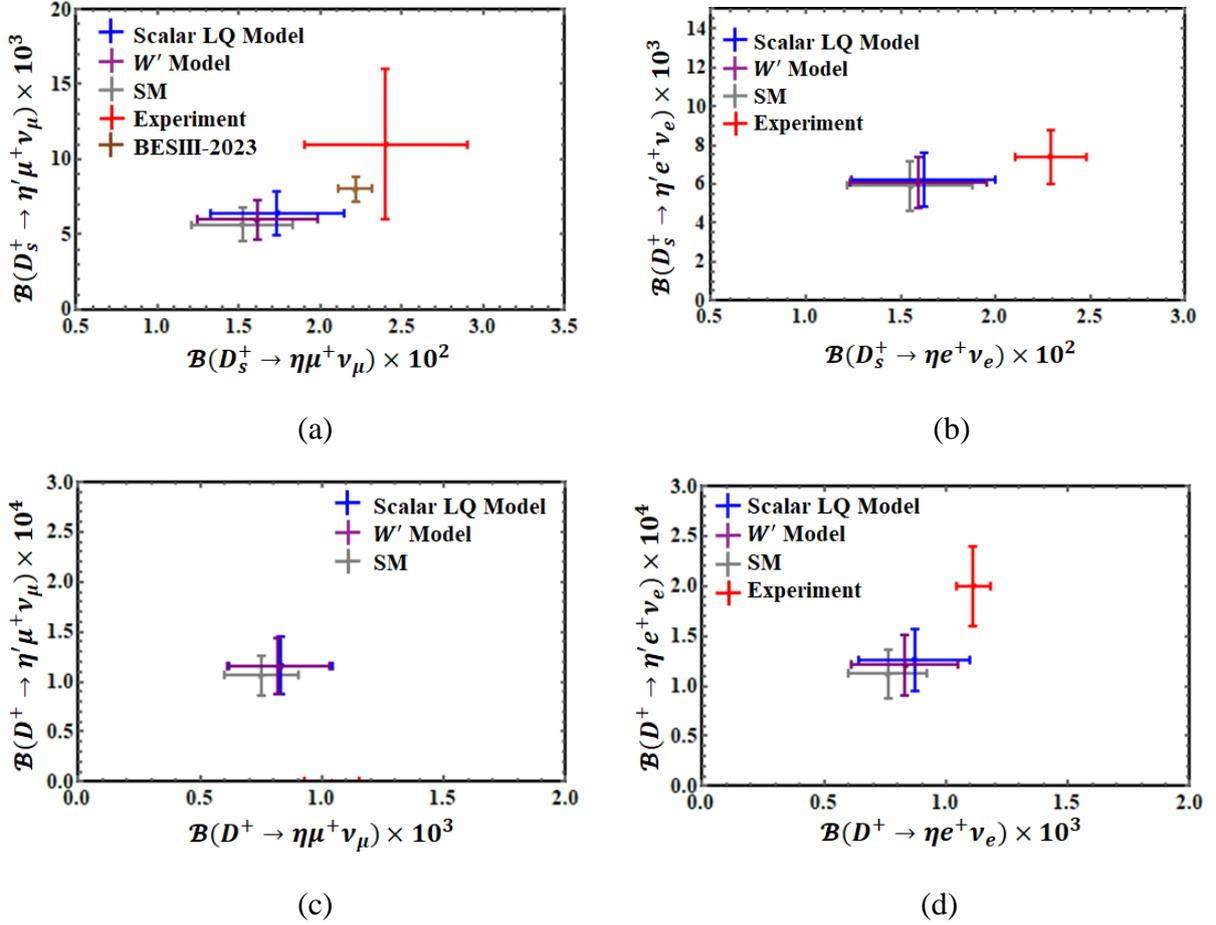

Fig 9. Comparison of the $W'$ model and scalar LQ model together with the experiment and SM in 2D $(\mathcal{B}(D_{(s)}^+ \to \eta l^+\nu_l), \mathcal{B}(D_{(s)}^+ \to \eta' l^+\nu_l))$ plane.

## 5. Conclusion

Although, recent experimental results on purely leptonic and semileptonic decays of $D$ mesons coincide with the theoretical predictions but a clear tension between theory and experiment has been observed on $D_{(s)}^+ \to \eta^{(\prime)}l^+\nu_l$ decays. So, some NP may exist in these channels. In this paper, we have investigated $D^+_{(s)} \to \eta^{(\prime)} l^+ \nu_l$ decays in $W'$ model and scalar LQ model to get possible NP footprint and we have also tried to explain the experimental results of branching fractions associated to these decays using these two NP models. In the Ref. [23], these decay modes have been studied with the $SU(3)$ flavour symmetry approach. Changing the hadronic form factors, they have constructed four different cases by considering $SU(3)$ flavour symmetry and symmetry breaking. In some cases, the predicted branching fraction results are within $2\sigma$ error range of the experimental results and some results are also consistent with experimental data. In the Refs. [18, 19], these channels have been investigated in model independent way. They have taken the NP contributions directly from the four-fermion operators (like vector, scalar and tensor operators) individually and the deviations from the SM have been shown graphically. In the present work, we have considered the contributions of above NP operators through the incorporations of new coupling parameters of the new particles like $W'$ boson and leptoquark. To structure these NP models, the coupling parameters are predicted from the recent experimental branching fraction results of semileptonic $D$ meson decays. Different decay observables like branching fraction, forward-backward asymmetry and lepton polarization asymmetry are studied within these NP models. Our predicted values of branching fractions in $W'$ model approach towards the experimental results and found around $1.5\sigma$ region. This model is unable to produce NP signal on forward-backward asymmetry and lepton polarization asymmetry because the NP dependent vector Wilson coefficient $C_{V_L}$ cancel out for these observables. Hence, $W'$ model is only suitable for studying branching fraction. Instead of considering the four-fermion operators individually (as done in Ref. [18, 19]), we have considered the combined impact of all these operators from their corresponding vector ($C_{V_L}$), scalar ($C_{S_L}$) and tensor ($C_T$) Wilson coefficients within the context of a scalar LQ model. Our predicted branching fraction results are found round the $1\sigma$ range of the experimental results excludeing $D_s^+ \to \eta e^+ \nu_e$ and $D^+ \to \eta' e^+ \nu_e$ decays and the effectiveness of our models is clearly observed through the comparison of our predicted results with the experimental results and SM predictions in 2D branching fraction plane. The key feature of the scalar LQ model is its capability of combining the impact of these operators together and due to the very small values of $C_T$, we have not observed any effect of scalar LQ on forward-backward asymmetry and lepton polarization asymmetry. Thus, from the above analysis it is cleared that, the scalar LQ model is more reliable to study $c \to (s, d)\bar{l}\nu_l$ transitions in the framework of NP. In fact, our predicted values in scalar LQ model, $\mathcal{B}(D_s^+ \to \eta \mu^+ \nu_\mu) = (1.73 \pm 0.41) \times 10^{-2}$ and $\mathcal{B}(D_s^+ \to \eta' \mu^+ \nu_\mu) = (6.42 \pm 1.47) \times 10^{-3}$ are consistent with the recent results by the BESIII collaboration [25], $(2.215 \pm 0.051 \pm 0.052) \times 10^{-2}$ and $(8.01 \pm 0.55 \pm 0.31) \times 10^{-3}$ respectively. The allowed coupling parametric space of the NP models with mass of these heavy bosons $\sim 1$ TeV reduces the deviation with the current experimental constraints (as shown in Table 7). We will be able to better comprehend the existence of NP with the help of the impending measurement of $D$ meson decays in the BESIII and Belle II experiments. We expect that our predicted values in scalar LQ model will be closer to the outcomes of upcoming experiments.

## Acknowledgement

S. Mahata and H. Mahapatra thank NIT Durgapur for providing fellowship for their research. M. Mandal acknowledges the Department of Science and Technology, Govt. of India for providing INSPIRE Fellowship through IF 200277.

## Appendix

Table A1. Values of the input parameters [24].

| Parameter | Value |
|---|---|
| $G_F$ | $1.1663787 \times 10^{-5}$ GeV$^{-2}$ |
| $V_{cs}$ | $0.987 \pm 0.011$ |
| $V_{cd}$ | $0.221 \pm 0.004$ |
| $m_D$ | $1869.66 \pm 0.05$ MeV |
| $m_{D_s}$ | $1968.35 \pm 0.07$ MeV |
| $\tau_D$ | $(1.033 \pm 0.005) \times 10^{-12}$ s |
| $\tau_{D_s}$ | $(5.04 \pm 0.04) \times 10^{-13}$ s |
| $m_\eta$ | $547.862 \pm 0.017$ MeV |
| $m_{\eta'}$ | $957.78 \pm 0.06$ MeV |
| $m_\mu$ | $0.1056583745$ GeV |
| $m_e$ | $0.51099895000$ MeV |
| $m_c$ | $1.27 \pm 0.02$ GeV |
| $m_s$ | $93.4^{+8.6}_{-3.4}$ MeV |
| $m_d$ | $4.67^{+0.48}_{-0.17}$ MeV |